\documentclass[prl,twocolumn]{revtex4}\usepackage{psfrag}
\usepackage{epsfig}
\usepackage{amsmath}
\usepackage{amssymb}
\newcommand{\bra}[1]{\ensuremath{\langle#1|}}
\newcommand{\ket}[1]{\ensuremath{|#1\rangle}}

\newcommand{\comm}[2]{\left [ #1, #2 \right]}
\newcommand{\be}{\begin{equation}}
\newcommand{\ee}{\end{equation}}
\newcommand{\avg}[1]{\ensuremath{\langle #1 \rangle}}
\newcommand{\id}{\openone}

\newcommand{\im}{\text{i}}

\newcommand{\ie}{{\it i.e.}}
\newcommand{\hc}{{\rm H.c.}}
\newcommand{\eg}{{\it e.g. }}

\newcommand{\etal}{{\it et al.}}

\newcommand{\eqcite}[1]{Eq.~\eqref{#1}}

\newcommand{\adop}{\hat a^{\dagger}}
\newcommand{\aop}{\hat a}
\newcommand{\bdop}{\hat b^{\dagger}}
\newcommand{\bop}{\hat b}

\newcommand{\xop}{\hat x}
\newcommand{\pop}{\hat p}

\newcommand{\sz}{\hat \sigma_{z}}
\newcommand{\sx}{\hat \sigma_{x}}

\newcommand{\spl}{\hat \sigma_{+}}
\newcommand{\smi}{\hat \sigma_{-}}

\begin{document}

\title{Quantum Magnetomechanics with Levitating Superconducting Microspheres}

\author{O. Romero-Isart}
\email{oriol.romero-isart@mpq.mpg.de}
\author{L. Clemente$^{1}$}
\author{C. Navau$^{2}$}
\author{A. Sanchez$^{2}$}
\author{J. I. Cirac$^1$}

\affiliation{$^1$Max-Planck-Institut f\"ur Quantenoptik,
Hans-Kopfermann-Strasse 1,
D-85748, Garching, Germany.}
\affiliation{$^2$Grup d'Electromagnetisme, Departament de F\'isica, Universitat Aut\`onoma de Barcelona, 08193 Bellaterra, Barcelona, Catalonia, Spain. }

\begin{abstract}
We show that by magnetically trapping a superconducting microsphere close to a quantum circuit, it is experimentally feasible to perform ground-state cooling and to prepare quantum superpositions of the center-of-mass motion of the microsphere. Due to the absence of clamping losses and time dependent electromagnetic fields, the mechanical motion of micrometer-sized metallic spheres in the Meissner state is predicted to be very well isolated from the environment. Hence, we propose to combine the technology of magnetic microtraps and superconducting qubits to bring relatively large objects to the quantum regime. 
\end{abstract}

\maketitle

The field of nano-mechanical resonators aims at cooling and controlling the  mechanical motion of massive objects in the quantum regime, and it is a subject of interest for both fundamental and applied science~\cite{ReviewOM}. The spectacular progress in this field has led to the achievement of ground-state cooling in a high-frequency micromechanical resonator using cryogenic refrigeration~\cite{O'Connell2010}, and in cavity electro-\cite{Teufel2011} and optomechanical~\cite{Chan2011} systems using sideband cooling techniques~\cite{Cooling}. The main challenge in these type of experiments is to overcome the heating and decoherence produced by clamping losses. These pose an even greater challenge for the experimental realization of some of the promised applications in the field, such as the preparation of quantum superposition states~\cite{Marshall2003}, which aim at the very fundamental goal of testing the validity of quantum mechanics when large masses are involved~\cite{Romero-Isart2011e}. Recently, a radical solution to this problem has been proposed, namely to unclamp the mechanical resonator and to use optical levitation instead~\cite{Romero-Isart2010b, Chang2010, Romero-Isart2011, Barker2010a, Singh2010b}. This is predicted to significantly improve the isolation of  the mechanical motion from the environment, even at room temperature. The archetypical scenario is a dielectric nanosphere trapped with optical tweezers inside a high-Finesse optical cavity. The dependence of the cavity resonance frequency on the center-of-mass position of the sphere yields an optomechanical coupling that can be employed to perform  ground-state cooling~\cite{Romero-Isart2010b, Chang2010}, as well as to prepare quantum superposition states~\cite{Romero-Isart2010b, Romero-Isart2011}. Remarkably, levitation is also the key ingredient in a recent proposal to prepare {\em large} spatial superpositions, namely of the order of the size of the nanosphere~\cite{Romero-Isart2011c, Romero-Isart2011e}.

\begin{figure}
\begin{center}
\includegraphics[width=0.8\linewidth]{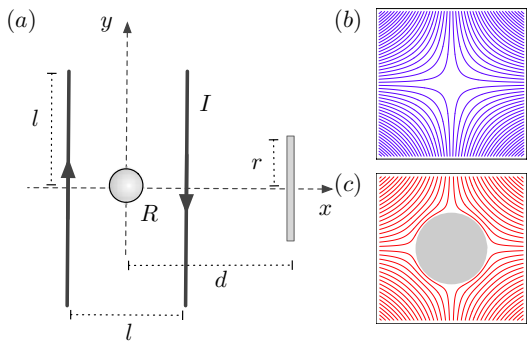}
\caption{(color online) (a) Scheme of the AHC with a pick-up coil. (b) Quadrupole field created by the AHC. (c) Field expelled by the sphere in the Meissner state. }
\label{Fig:2DScheme}
\end{center}
\end{figure}

A common feature in the vast range of optomechanical systems~\cite{ReviewOM}, including optically levitating mechanical oscillators~\cite{Romero-Isart2010b, Chang2010, Barker2010a, Romero-Isart2011,Singh2010b}, is that photons are used to cool and manipulate the mechanical motion. This introduces two important limitations, which are, indeed, the main sources of decoherence for optically levitating objects: (i) scattering of photons, that produces position localization decoherence~\cite{Romero-Isart2010b, Chang2010,Romero-Isart2011c}, and (ii) absorption of photons,  that increases the bulk temperature of the object, and, consequently, also the decoherence due to emission of black body radiation~\cite{Chang2010, Romero-Isart2011c, Romero-Isart2011e, Romero-Isart2010b}. In this Letter, we propose a levitating mechanical resonator experimental setup which does not employ photons but magnetostatic fields. Therefore, it does not only avoid clamping losses due to levitation, but it also circumvents limitations (i) and (ii) due to the absence of photons when trapping, cooling, and coherently manipulating the object. Hence, this mechanical oscillator is predicted to be very well isolated from the environment. This leads to very large mechanical quality factors and long coherences times (due to negligible black-body radiation), the latter being specially relevant in time-of-flight experiments~\cite{Romero-Isart2011c, Romero-Isart2011e}. The proposal consists in magnetically trapping a superconducting microsphere in the proximity of a quantum circuit (\eg LC oscillator or flux qubit)~\cite{Makhlin2001}. The magnetic field employed in the magnetic trap is expelled~(see Fig.~\ref{Fig:2DScheme}), due to the Meissner effect, from the superconducting sphere. The flux passing through the pick-up coil of the quantum circuit  depends on the center-of-mass position of the sphere. This leads to a significant quantum magnetomechanical coupling with the center-of-mass motion. Here we show that ground-state cooling and the preparation of quantum superpositions can be achieved, within the same experimental setup and with feasible parameters, for superconducting spheres (\eg Pb) in the $\mu$m regime with masses of $\sim 10^{14}$ amu. 

More specifically, we consider a superconducting microsphere of radius $R$ and mass $M$ that is in the Meissner state, cooled below a certain temperature $T_{\text{C}}$. The penetration length $\lambda$ and coherence distance $\xi$ are such that $R \gg \lambda, \xi$. In this regime, one can approximate that the magnetic induction ${\bf B}$ is zero in the whole interior of the superconductor. The microsphere is confined into a 3D harmonic potential created by a magnetic microtrap~\cite{MagTraps}. While the proposal does not rely on the particular trapping scheme, we consider here a quadrupole trap created by two circular coils of radius $l \gg R$ in an anti-Helmholtz configuration (AHC), \ie~the coils are coaxials, separated by a distance $l$, and with opposite current intensity $I$, see Fig~\ref{Fig:2DScheme}. This trap creates an harmonic potential of the form $\hat V_\text{trap}= M [ \omega^{2}_\text{t} \xop^{2} + \omega^{2}_{\perp} (\hat y^{2} + \hat z^{2})]/2$. The trapping frequency $\omega_\text{t}$ can be obtained analytically using the image method~\cite{Lin2006} due to the cylindrical symmetry of the system, see the Supplemental Material (SM). The transverse frequency can be derived by replacing the sphere by an effective magnetic moment. Their expressions are given by~$\omega_\text{t}\simeq 1.05 \sqrt{(\mu_{0}/\rho)} I/l^{2}$ and $\omega_{\perp}=\omega_\text{t}/2$. $\mu_{0}$ is the vacuum permeability and $\rho$ the density of the microsphere which is assumed to be homogeneous. An important remark is that the field at any point of the sphere is required to be smaller than the critical field $B_{\text{C}}$ in order to allow superconductivity; this yields an upper bound on the radius of the sphere $R < R_{\text{max}} \simeq 0.98 B_{\text{C}}/(\omega_\text{t}\sqrt{\mu_{0} \rho})$. 

At some distance $d$ from the center of the trap and coaxially with the AHC (Fig~\ref{Fig:2DScheme}) 
a pick-up coil of radius $r$ is placed (the perpendicular configuration might also be considered, see SM). The pick-up coil is connected to a quantum LC oscillator or to a flux qubit~\cite{Makhlin2001} placed outside the magnetic field of the trap. As shown below, the quantum magnetomechanical coupling for both cases depends on the dimensionless parameter $\eta \equiv x_\text{zp}  \Phi_\text{ext}'(0) / \Phi_{0}$, where $x_\text{zp}=\sqrt{\hbar/(2M\omega_{\text{t}})}$ is the mechanical zero point motion, $\Phi_{0}=\pi \hbar/e$ is the flux quantum, and $\Phi_\text{ext}'(0)$ is the derivative with respect to the axial motion of the center-of-mass evaluated at its equilibrium position of the flux threading the pick-up coil. $\Phi_\text{ext}'(0)$ can also be evaluated analytically using the image method~\cite{Lin2006} and its approximated expression for $R \ll l,r$ is given by  $\Phi_\text{ext}'(0) \approx 2.7 \mu_{0} (I/l^{2})  R^{3} r^{2}/ ( d^{2}  + r^{2})^{3/2}$ (see SM for further details).  

For an LC oscillator, consisting of an inductor $L$ connected to a capacitor $C$, the Hamiltonian is given by $\hat H_\text{LC}= [\hat \Phi - \Phi_\text{ext} (\xop)]^{2}/(2L) + \hat Q^{2}/(2C) $, with $[\hat \Phi,\hat Q]= \im \hbar$. By expanding $\Phi_\text{ext}(\xop)$ linearly in $\xop$, one obtains the linear Hamiltonian $\hat H_\text{LC}= \hbar\omega_\text{LC} \adop \aop + \hbar g_{\text{LC}} (\adop + \aop) (\bdop + \bop)$. We used $\xop = x_\text{zp} (\bdop + \bop)$ and $\hat \Phi = \Phi_\text{zp} (\adop + \aop)$, where $\Phi_\text{zp}=\sqrt{\hbar/(2 C \omega_\text{LC})}$ is the zero point flux, $\omega_\text{LC}=1/\sqrt{LC}$ is the LC resonator frequency, and $g_\text{LC}= \epsilon_\text{LC}\eta$, with $\epsilon_\text{LC}= \Phi_{0} \Phi_\text{zp}/(\hbar L)$, is the magnetomechanical coupling rate. Alternatively, one can consider a three junction flux qubit~\cite{Mooij1999,Makhlin2001}.  The two qubit states correspond to persistent currents of amplitude $I_\text{p}$ flowing clockwise or counterclockwise in a loop of radius $r$.  When the qubit is operated in the vicinity of the degeneracy point $f(\Phi_\text{ext})Ê\equiv \Phi_\text{ext}/\Phi_{0} - 1/2 \approx 0$, the Hamiltonian of the qubit in the basis of the persistent current states reads $\hat H_\text{s}=-\hbar \tilde \epsilon \sz/2 - \hbar \Delta \sx/2$. Here $\hat \sigma_{i}$ ($i=x,y,z$) are the usual Pauli matrices, $ \tilde \epsilon= \nu f(\Phi_\text{ext}) /\hbar $ is the bias (where $\nu=2 \Phi_{0} I_\text{p} $), and $\Delta$ is the tunneling amplitude. Thus, by expanding $\tilde \epsilon(\xop) \approx \tilde \epsilon(0) + \tilde \epsilon'(0) \xop$ to first order in $\xop$, one obtains the quantum magnetomechanical Hamiltonian (the harmonic oscillator energy is also included),
\be \label{eq:HMM}
\frac{\hat H_{\text{MM}}}{\hbar}=  \omega_{\text{t}} \bdop \bop - \frac{\epsilon}{2} \sz -  \frac{\Delta}{2} \sx -  g_{0} \hat \sigma_{z} (\bdop + \bop).
\ee
Here, $\epsilon=\tilde \epsilon(0)$, and $g_{0}=\nu \eta$ is the non-linear magnetomechanical coupling. The energy levels of the qubit are separated by a frequency $\omega_\text{s}= \sqrt{\epsilon^{2} + \Delta^{2}}$. It is remarkable that with typical numbers, see below, $\nu$ is only two or three orders of magnetic lower than  $\epsilon_\text{LC}$ since the coupling to the flux qubit provides a non-linearity to the mechanical motion. For this reason, we concentrate hereafter on the coupling with the flux qubit in order to discuss ground-state cooling and the preparation of quantum superpositions.  However, due to its larger coupling, the LC oscillator would be definitely more appropriate for performing cooling or any other Gaussian dynamics (\eg preparation of squeezed states or entanglement and teleportation in the case of several spheres). In this case a parametric driving in the magnetomechanical coupling (\eg with a time-dependent inductance) would be required. We remark that while we do not consider it here, coupling to the transversal motion also leads to a significant magnetomechanical coupling.

Decoherence in superconducting qubits is modeled with a master equation containing the following Lindbland terms~\cite{Makhlin2001} (written in the eigenbasis of the qubit): spontaneous emission $\mathcal{L}_{0}[\hat \rho]= \Gamma_{0}(2 \smi \hat \rho \spl- \hat \rho \spl \smi - \spl \smi \hat \rho)/2$, and pure dephasing $\Gamma_{\varphi}$, $\mathcal{L}_{\varphi}[\hat \rho]= \Gamma_{\varphi} (\sz \hat \rho \sz - \hat \rho)/2$.  The decoherence rates are related to the relaxation time $T_{1}=1/ \Gamma_{0}$, which is the time required for the qubit to relax from the excited state to the ground state, and to the dephasing time $T_{2}=(\Gamma_{0}/2 +   \Gamma_{\varphi})^{-1}$, which is the time over which the phase difference  between two eigenstates becomes randomized. 

The distinctive feature of this proposal is that the decoherence in the mechanical oscillator, when accounting for known sources, is predicted to be very small since both (i) clamping losses and (ii) scattering of photons are absent. Similarly as in the proposal of optically levitating dielectrics~\cite{Romero-Isart2010b, Chang2010, Romero-Isart2011}, other sources are also negligible: (iii) damping created by the background gas yields a mechanical quality factor that can reach extremely high values at sufficiently low pressure, this is given by~\cite{Romero-Isart2010b,Chang2010} $Q_\text{air}=\omega_\text{t}/\gamma_\text{air} \gtrsim10^{11}$,  where $\gamma_\text{air}= 16 P /(\pi \bar v R \rho)$, $P\sim 10^{-10}$ Torr is the environmental pressure, and $\bar v$ the thermal mean velocity of the air molecules; (iv) decoherence due to blackbody radiation~\cite{Chang2010,Romero-Isart2011c,Romero-Isart2011e} yields rates at the Hz regime due to the cryogenic bulk temperatures of the metallic superconducting microsphere (in optical levitation the bulk temperature of the object is heated due to light absorption); (v) internal vibrational modes are decoupled (due to their higher frequency) to the center-of-mass motion for micrometer-sized objects, see~\cite{Romero-Isart2011} for a detailed analysis based on quantum elasticity. Other sources of decoherence particular to this proposal are: (vi) damping due to hystereses losses in the superconducting coils yield $QÊ\gg 10^{10}$ (specially for small fluctuations in the position, see SM for details); (vi) fluctuations in the trap frequency leads to decoherence with a rate given by~\cite{Gehm1998} $\Gamma_\omega=\pi \omega_\text{t}^{2} S_{\omega} (2 \omega_\text{t})/16$, where $S_{\omega} (2 \omega_\text{t}) $ is the one-sided power spectrum of the fractional fluctuation in the resonance frequency (see SM for details). $\Gamma_{\omega} \sim \text{Hz}$ can be obtained for $\sqrt{S_{\omega} (2 \omega_\text{t}) }= 10^{-5}/\sqrt{\text{Hz}}$; (vii) fluctuations in the trap center also lead to decoherence with the rate~\cite{Gehm1998}  $\Gamma_{x}=\pi \omega^{2}_\text{t}  S_{x} (\omega_\text{t})/4x_\text{zp}^2 $, where in this case $S_{x}$ is the one-sided power spectrum of the position fluctuations. $\Gamma_{x} \sim \text{Hz}$ can be obtained for position stability $\sqrt{S_{x} (\omega_\text{t})}/x_\text{zp} \sim 10^{-4}/\sqrt{\text{Hz}}$; (viii) within trapping frequencies in the kHz regime, which are much smaller than the energy gap ($100$ GHz), the superconductor can be considered to act instantaneously to external fields; (ix) and finally, we remark that the flux qubit present in the setup can always be decoupled from the center-of-mass motion by switching off the driving such the coupling is off-resonant (see below). Thus, at sufficiently low pressure and for superconductors in the Meissner state, the center-of-mass of micrometer-sized metallic spheres is (according to the sources of decoherence that we have considered) effectively decoupled from the environment. This assumes very stable traps (both in the frequency and in the equilibrium position). In the actual experiment, other uncontrolled sources of decoherence might be relevant, for instance, coupling to bond paramagnetic centers on silicon surface have been experimentally observed at distances of few micrometers~\cite{Vinante2011}.

\begin{figure}
\begin{center}
\includegraphics[width=\linewidth]{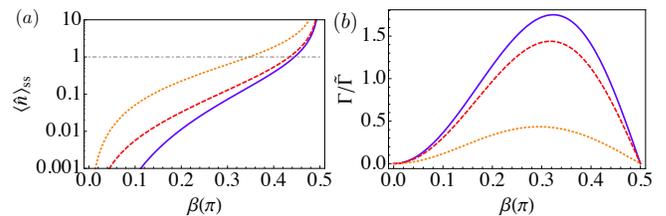}
\caption{(color online)  (a) The steady state phonon number occupation $\avg{\hat n}_\text{ss}$ (assuming $\Gamma_\text{ext}=0$) and (b) the cooling rate $\Gamma$ over  $\tilde \Gamma= \tilde g^{2} \cos^{2} (\alpha)/\Gamma_{0}$ are plotted as a function of $\beta$ for $\Gamma_\varphi/\Gamma_{0}=0$ (solid blue line), $\Gamma_\varphi/\Gamma_{0}=0.1$ (dashed red line), and $\Gamma_\varphi/\Gamma_{0}=1$ (dotted orange line). }
\label{Fig:Cooling}
\end{center}
\end{figure}

We now focus on ground-state cooling of the mechanical motion. To this end, a resonant coupling between the qubit and the mechanical resonator is required~\cite{Armour2002,Rabl2009}. Note however that the energy splitting of the qubit $\omega_{\text{s}}$ (in the GHz regime) is much larger the mechanical frequency $\omega_{\text{t}}$ (in the kHz-MHz regime) of the oscillator. This scenario has been studied both experimentally and theoretically in~\cite{Ilichev2003}, where a flux qubit has been coherently coupled to a slow LC oscillator.  This is achieved by driving the flux qubit with an applied ac flux with frequency $\omega_\text{d}$ (in the GHz regime) and amplitude $\Omega$ (in the kHz-MHz regime). In this case, the total Hamiltonian reads $\hat H_\text{t}=\hat H_\text{MM} + \hat H_{\text{drive}}$, where $\hat H_{\text{drive}}=  \hbar \Omega \cos(\omega_{\text{d}} t)  \sz $. Recall that the dynamics of the qubit and the mechanical oscillator (which is decoupled from the environment) is given by the master equation
$\dot \rho = - \im [\hat H'_\text{t},\hat \rho]/\hbar + \mathcal{L}_{0}[\hat \rho] + \mathcal{L}_{\varphi}[\hat \rho]$,
where $\hat H_\text{t}'$ is the total Hamiltonian written in the eigenbasis of the qubit.  By moving to a rotating frame of the qubit at frequency $\omega_\text{d}$, performing a rotating wave approximation (RWA) (valid provided that $\omega_\text{d} \sim \omega_\text{s} \gg \omega_\text{t}, g_{0}$), transforming to the diagonal basis of the qubit, moving to the interaction picture, and performing a second RWA (assuming $\omega_{t} \approx \tilde \omega_\text{s} \gg g_{0}, \Gamma_{0},\Gamma_\varphi$, where $\tilde \omega_\text{s} = (\delta \omega^{2} + \tilde \Omega^{2})^{1/2}$,  $\delta \omega = \omega_\text{d} - \omega_\text{s}$, $\tilde \Omega= \Omega \sin \alpha$, and $\tan \alpha = \Delta/\epsilon$), one arrives at
\be
\dot \rho = \frac{\im}{\hbar} \comm{\tilde g (\smi \bdop + \hc)}{\hat \rho} + \mathcal{L}_\Gamma[\hat \rho].
\ee
The effective magnetomechanical coupling is given by $\tilde g = g_{0} \cos \alpha \sin \beta$, where $\tan \beta= \tilde \Omega/\deltaÊ\omega$. The dissipation of the qubit is given by $\mathcal{L}_\Gamma[\hat \rho]$ and contains both dephasing  $\Gamma^{\star}_{\varphi} (\sz \hat \rho \sz - \hat \rho)/2$, and transitions in both directions $\Gamma_{\downarrow (\uparrow) }(2 \hat \sigma_{\mp} \hat \rho \hat \sigma_{\pm}- \hat \rho \hat \sigma_{\pm} \hat \sigma_{\mp} - \hat \sigma_{\pm} \hat \sigma_{\mp} \hat \rho)/2$. The rates are given by $\Gamma^{\star}_{\varphi}= \Gamma_{\varphi} \cos^{2} \beta+ \Gamma_{0} \sin^{2}(\beta) /2$, and $\Gamma_{\downarrow (\uparrow)}= \Gamma_{\varphi}\sin^{2}(\beta)+ \Gamma_{0} (1\pm\cos \beta)^{2}/2$. Typically $\tilde g \ll \Gamma_{\downarrow (\uparrow)}, \Gamma^{\star}_{\varphi}$, which allows to adiabatically eliminate the qubit~\cite{Cirac1992}. This leads to an effective master equation describing the mechanical oscillator density matrix that can be used to obtain a dynamical equation for the mean phonon number occupation $\hat n \equiv \bdop \bop$, namely
$\avg{\dot n} = - \Gamma \avg{\hat n} + A_{+}+\Gamma_\text{ext}$. The cooling rate is defined as  $\Gamma=A_{-}-A_{+}$, where
\be
A_{\pm} = \frac{2 \tilde g^{2}}{  \left ( 2 \Gamma^{\star}_{\varphi} + \Gamma_{\uparrow} + \Gamma_{\downarrow} \right)} \left(1 \mp \frac{\Gamma_{\downarrow} - \Gamma_{\uparrow}}{\Gamma_{\downarrow} + \Gamma_{\uparrow}} \right). 
\ee
The rate $\Gamma_\text{ext}$ takes into account external sources of heating in the mechanical oscillator (\eg due to trap fluctuations) which are assumed to be much smaller than $\Gamma$.  
The final phonon number occupation is given by $\avg{\hat n}_\text{ss}=(A_{+}+\Gamma_\text{ext})/\Gamma$. In Fig.~\ref{Fig:Cooling} it is shown that ground-state cooling can be achieved within a wide range of $\beta$ with cooling rates of the order of $\sim \tilde g^{2} \cos^{2} (\alpha)/\Gamma_{0}$ ($\alpha < \pi/2$ is also required). 

\begin{figure}
\begin{center}
\includegraphics[width=\linewidth]{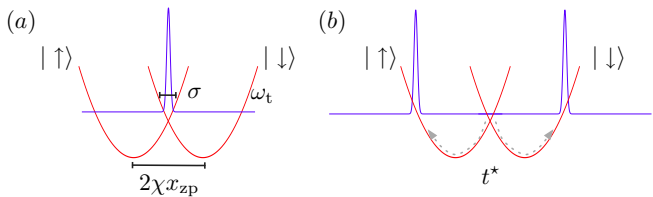}
\caption{(color on-line) Scheme of the protocol to prepare quantum superpositions of a mechanical oscillator using the parametric coupling to a qubit. (a) The state $\ket{+} \ket{0}$ that is prepared $t=0$ and recovered at $t=2t^{\star}$. (b) The superposition state \eqcite{eq:state} that is created at $t=t^{\star}$. }
\label{Fig:Superposition}
\end{center}
\end{figure}

The non-linearity given by the presence of the qubit can be readily employed to prepare quantum superpositions of the mechanical oscillator~\cite{Armour2002}. First, we rotate the Hamiltonian \eqcite{eq:HMM} to the eigenbasis of the qubit, we move to the interaction picture (using as a free part the qubit energy term), and perform the RWA (valid provided $\omega_\text{s} \gg g_{0}$). We arrive at $\hat H'_\text{MM} = \hat H_{\text{m}}+\hbar \omega_\text{s} \sz - g \sz \xop/x_\text{zp} $, where $g=g_{0}\cos \alpha$ and $\hat H_\text{m}=\pop^{2}/(2M) + M \omega^{2}_\text{t} \xop^{2}/2$. This can be re-written as 
$\hat H'_{\text{MM}}=  \hbar \omega_{\text{s}}/2 \sz + \hat T^{\dagger}( \chi \sz ) \hat H_\text{m} \hat T (  \chi  \sz )$, 
where $\hat T(a) = \exp[-\im \pop a x_\text{zp} /\hbar]$ is the translational operator (such that $\hat T^{\dagger}(a) \xop \hat T(a)=\xop+a x_\text{zp}$), $\chi \equiv 2g /\omega_{\text{t}}$ is a dimensionless parameter, and the constant term $-\hbar g^{2}/\omega_{\text{t}} \id $ has been dropped out.  The Hamiltonian written in this form points out the key property used in the protocol, that is, that the center of the harmonic trap depends on the spin state of the qubit, see Fig.~\ref{Fig:Superposition}(a). Hence, by initially preparing the joint system into the state $\ket{\Psi_{0}} = \ket{+,0} =  \left(\ket{\uparrow,0} + \ket{\downarrow,0} \right)/\sqrt{2}$, that is, the qubit in the superposition state $\ket{+}= (\ket{\uparrow} + \ket{\downarrow})/\sqrt{2}$ and the mechanical oscillator is in a pure state $\ket{0}$ with $\avg{\xop^{2}}=\sigma^{2}$, the joint state evolves after time $t_{\star}=\pi/\omega_\text{t}$ into the entangled state
\be \label{eq:state}
\ket{\Psi_\text{s}} =  \frac{1}{\sqrt{2}} \left[ \hat T(-2 \chi) \ket{\uparrow,0} + \hat T(2\chi) \ket{\downarrow,0} \right].
\ee
Hence, the mechanical oscillator is in a spatial quantum superposition state (Fig.~\ref{Fig:Superposition}(b)) where the wave packets are separated by a distance $l_\text{s} =4 \chi x_\text{zp}= 8 x_\text{zp} g/\omega_\text{t}$. The overlap is given by $\bra{0} \hat T^{\dagger}(-2\chi) \hat T(2\chi)\ket{0}=\exp \left[ - l^{2}_\text{s} /(8 \sigma^{2}) \right]$, and thus $8 \sigma^{2} < l^{2}_\text{s}$ is required. Due to levitation, this challenging condition~\cite{Armour2002} can be guaranteed by initially squeezing the ground state of the mechanical oscillator, see SM. The superposition can be probed by performing tomography of the qubit during the evolution of the joint state within a time window $t \in [0,2 t^{\star}]$, such that the collapse at $t=t^{\star}$(due to the entangled state \eqcite{eq:state}) and revival at $t=2 t^{\star}$ (due to the product state) of the purity of the qubit state can be observed. Decoherence in the qubit can be neglected provided $2 t^{\star} \ll T_{2}$. Note that for long coherence times in the qubit, the superposition size could be increased by performing a spin flip after each $t_{\star}$ evolution, or alternatively, by opening the trap to a frequency $\omega' \ll \omega_\text{t}$.

Regarding the experimental feasibility of this proposal, in the following we choose some set of parameters that fulfill the conditions that are required. We consider a microsphere made of Pb, with density $\rho=11360$ Kg/$\text{m}^{3}$, penetration depth $\lambda=30.5$~nm and  coherence length $\xi=96$~nm (evaluated at $T=0$), critical temperature $T_\text{C}=7.2$~K, and critical field $B_\text{C}=0.08$~T. The magnetic trap is made with AH coils of radius $l=25$~$\mu$m and current $I=10$~A (assuming materials with critical current $J_c=7 \times 10^{11}$A/m$^2$~\cite{dikovsky2009}) such that $\omega_\text{t} \approx 2 \pi  \times 28 $~kHz and $R_\text{max}=3.7$~$\mu$m. Thus, we consider a sphere of radius $R=2$~$\mu$m. The pick-up coil has a radius $r = 24.5$~$\mu$m and is placed at $d=17.5$ $\mu$m (outside the AHC). This yields $\eta= 1.3 \times 10^{-7}$. For the flux qubit we use~\cite{Makhlin2001} $\nu=\Delta=2 \pi \times 10$ GHz and $T_{1}=T_{2}=10$~$\mu$s, and hence we obtain $g_{0}= 2 \pi \times 1.3 $~kHz, $\Gamma_{0}= 2 \pi \times 16 $~kHz and $\Gamma_{\varphi}= \Gamma_{0}/2 $. For an LC oscillator with $C=1$ pF and $L=0.1$ nH~\cite{Makhlin2001}, the magnetomechanical coupling is nearly two orders of magnitude larger, namely $g_\text{LC}=2 \pi \times 93 $ kHz. This set of parameters allows for ground-state cooling and the preparation of quantum superposition states using the squeezing of the ground-state wavefunction.

We have shown that micrometer-sized superconducting metallic spheres containing $\sim 10^{14}$ atoms can be cooled down to the ground state and prepared into superposition states. From a broader perspective, while cavity optomechanics is using the technology developed for laser trapping, cooling, and manipulation of atoms and ions, here we propose to merge the  technology of the growing fields of magnetic trapping of atoms and superconducting qubits to bring massive objects into the quantum regime. The combined properties of levitation, low bulk temperatures, and large masses, makes this setup ideally suited to design and implement protocols where the objects is released from the trap in order to expand the wavefunction. This can be used to create large superpositions~\cite{Romero-Isart2011c} in order to test fundamental questions~\cite{Romero-Isart2011e} and to design ultra-high sensitive detectors.

We acknowledge funding from EU project MALICIA, DFG SFB 631, and Spanish Consolider Project NANOSELECT (CSD2007-00041). {\em Note added}: We have become aware of a recent proposal of a levitated magnetomechanical system by M.~Cirio, J.~Twamley, and G.~K.~Brennen, arXiv:1112.52086.

\end{document}